\documentstyle[epsf,psfig,aps]{revtex}
\begin{document}
\twocolumn[\hsize\textwidth\columnwidth\hsize\csname
@twocolumnfalse\endcsname
\title{Validity of Feynman's prescription of disregarding the Pauli
principle\\ in intermediate states}

\author{F.A.B. Coutinho$^a$, Y. Nogami$^{b,c}$ and Lauro Tomio$^c $}

\address{
$^a$ Faculdade de Medicina, Universidade de S\~ao Paulo, \\
Av. Dr. Arnaldo, 455, 01246-903, S\~ao Paulo, Brazil \\ 
$^b$ Department of Physics and Astronomy, McMaster University,\\
Hamilton, Ontario, Canada L8S 4M1 \\ 
$^c$ Instituto de F\'{\i}sica Te\'orica, Universidade
Estadual Paulista, \\
Rua Pamplona, 145, 01405-900, S\~ao Paulo, Brazil }

\date{\today}
\maketitle

\begin{abstract}
Regarding the Pauli principle in quantum field theory and in
many-body quantum mechanics, Feynman advocated that Pauli's
exclusion principle can be completely ignored in intermediate 
states of perturbation theory. He observed that all virtual 
processes (of the same order) that violate the Pauli principle 
cancel out. Feynman accordingly introduced a prescription, which 
is to disregard the Pauli principle in all intermediate processes.
This ingeneous trick is of crucial importance in the Feynman 
diagram technique. We show, however, an example in which Feynman's 
prescription fails. This casts doubts on the general validity of 
Feynman's prescription. 
\end{abstract}
\vskip2pc]
\vskip 1cm

\section{Introduction}
In his space-time approach to quantum electrodynamics, Feynman
advocated: ``It is obviously simpler to disregard the exclusion
principle completely in the intermediate states" \cite{1}. He
examined processes involving several particles and observed that
all virtual processes that violate Pauli's exclusion principle
(formally) cancel out. It is understood that all virtual
processes of the same orders are taken into account. On the basis 
of this observation Feynman introduced a prescription that is 
to disregard the Pauli principle in all intermediate states. This 
ingenious trick was crucial in accomplishing the enormous 
simplification and transparency of perturbation theory. 
For example, the vacuum polarization can be related to Feynman
diagrams with an electron loop or loops. Although the process
represented by a loop diagram may (at least partially) violate
Pauli's exclusion principle, no restriction needs to be imposed
on integrations with respect to associated momentum variables.
Feynman's prescription is also often used in perturbation 
calculations for many-body systems in quantum mechanics.

Various aspects of Feynman's prescription have been discussed 
by several authors \cite{1a}. There are some intriguing 
implications regarding the meson effects in nuclei or nuclear 
matter. Feynman's prescription is instrumental in proving
Goldstone's theorem for many-body systems. We are not going to 
review these topics in this paper but we emphasize that no 
suspicion seems to have ever been raised in the literature
against the validity of Feynman's prescription.

The purpose of this paper we present an example that casts
doubt about the general validity of Feynman's prescription. The 
example is concerned with the second order energy shift of a
relativistic bound system. We consider a model that consists of
a particle bound in a given potential. The wave function of the
particle is subject to the Dirac equation with the given binding
potential. In addition to the bound particle, there is a vacuum
background. It is understood that the vacuum background is an
integral part of the bound system. When an external perturbation
is applied, the energy of the system is shifted. We calculate
the energy shift in second order perturbation theory. We are
particularly interested in the vacuum effect to which the Pauli
principle is relevant.

We consider two methods, I and II, for calculating the energy
shift. In method I we take account of the Pauli principle
whenever it is applicable. In method II we disregard the Pauli
principle altogether. We confirm that these two methods formally 
agree. This illustrates Feynman's prescription. When methods
I and II are explicitly worked out for the example, however, 
the results of the two methods turn out to disagree with each
other. We analyze the intriguing mechanism of this discrepancy.

In Sec. II we set up the model and illustrate Feynman's 
prescription. In Sec. III we make the model more explicit. We 
consider a charged particle that is bound in an infinite 
square-well potential of the Lorentz scalar type. This is a 
one-dimensional version of the ``bag model". For the external 
perturbation we assume a homogeneous electric field. Then the 
second order energy shift is related to the electric polarizability 
of the system. We carry out the calculations of methods I and II. 
The two methods result in different energy shifts (and hence
different values of the electric polarizability). We analyze the
source of the discrepancy. In Sec. IV we confirm the result of
method II by repeating the calculation by using the Dalgarno-Lewis 
(DL) method \cite{2,3,4}. A summary and discussions are given in 
Sec. V. Some details concerning the series that appear in 
method II are relegated to the Appendix.

\section{Feynman's prescription}
As a way of setting up notation, let us start with a problem of
single-particle quantum mechanics. Let the Hamiltonian of the
model be
\begin{equation}
H = H_0 + V \, , 
\label{1}
\end{equation}
where $H_0$ is the Dirac Hamiltonian with a binding potential
and $V$ is the external perturbation. (Imagine something like a
hydrogen atom, with Hamiltonian $H_0$, placed in a weak external
electric field $V$. Assume that the proton is merely a source of
the Coulomb potential that binds the electron of the atom.) We
take $H_0$ as the unperturbed Hamiltonian and treat $V$ by
perturbation theory. It is understood that the solutions of the
Dirac equation with $H_0$ are known for all stationary states,
\begin{equation}
H_0|i\rangle  = \epsilon_{i}|i\rangle \,  ,\hspace{0.3in}
H_0|-j\rangle  = \epsilon_{-j}|-j\rangle \, ,
\label{2}
\end{equation}
where $i=1,2, \cdots$ and $-j=-1,-2, \cdots$ . The $|i\rangle$'s 
are positive energy states with $\epsilon_i > 0$ and 
$|-j\rangle $'s are negative energy states with
$\epsilon_{-j} <0$. In particular $|1\rangle $ is the lowest
positive energy state (like the 1$S$-state of the hydrogen atom). 
We are assuming that the eigenvalues are all discrete. (It is 
not difficult to include continuum states; see Sec. V.)  The 
$|i\rangle $'s and $|-j\rangle $'s form a complete orthonormal 
basis set. Figure 1 schematically shows the unperturbed energy 
spectrum.

\begin{figure}[htb]
\vskip -0.7cm
{\psfig{figure=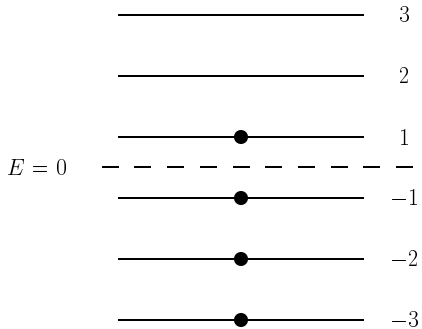,width=3in}}
\vspace{-5.5cm}
\caption{
Unperturbed energy levels defined by Eq. (\ref{2}). The dots
indicate occupied states.
}\label{fig1}
\end{figure}

For the unperturbed state let us take $|1\rangle$, the state of
the lowest positive energy \cite{5}. In single-particle quantum
mechanics we do not consider the vacuum background. In other words, 
we assume that the negative energy states are all empty. In
this sense let us momentarily ignore the dots of the states of the 
$|-j \rangle $'s in Fig. 1. Then the energy shift $W_{\rm QM}$ of 
state $|1\rangle $ caused by perturbation $V$ is given by
\begin{equation}
W_{\rm QM} = \sum_{i\neq 1} \frac{|V_{i,1}|^2}{\epsilon_1
-\epsilon_i} + \sum_{j} \frac{|V_{-j,1}|^2}{\epsilon_1
-\epsilon_{-j}}\, ,
\label{3}
\end{equation}
where $V_{i,1}\equiv \langle i|V|1\rangle$ and $V_{-j,1}\equiv
\langle -j|V|1\rangle$.  Suffix QM refers to single-particle
quantum mechanics. The summation for $i$ ($j$) is for the
positive (negative) energy intermediate states.

Let us now take account of the vacuum background that accompanies
the particle bound in $|1\rangle$. Let us examine the vacuum from 
the point of view of Dirac's hole theory. The vacuum consists of 
an infinite number of particles that occupy all of the negative 
energy states as indicated with dots in Fig. 1. Before the external 
perturbation $V$ is applied, the negative energy states are 
eigenstates of $H_0$. The energy of the unperturbed system, 
including the vacuum background, is
\begin{equation}
E = \epsilon _1 + \sum_{j} \epsilon _{-j} \, .
\label{4}
\end{equation}
The sum over the negative energy states is the energy of the
vacuum. This vacuum is different from the ``free vacuum''. The 
$\epsilon _j$'s are eigenvalues of $H_0$ that contains the binding
interaction. The summation of Eq. (\ref{4}) diverges but this is 
not a serious problem. If we subtract the energy of the free 
vacuum, the summation will converge but let us not delve into this 
aspect. We are only interested in the energy shift rather than the 
energy itself. All $\epsilon$'s and hence the total energy $E$ are 
shifted by external perturbation $V$. Let us examine two methods, 
I and II, for calculating the energy shift.

{\em Method I}. In this method we take account of the Pauli
principle in intermediate states. Energy $\epsilon _1$ is
shifted by
\begin{equation}
W_1 = \sum_{i\neq 1} \frac{|V_{i,1}|^2}{\epsilon_1 -\epsilon_i} \, ,
\label{5}
\end{equation}
where the Pauli principle excludes the negative energy states as
intermediate states. The difference between $W_1$ and $W_{\rm
QM}$ is due to the presence of the vacuum background in the
former. On the other hand the vacuum energy itself also shifts
by
\begin{equation}
W_{\rm vac} = \sum_{j}W_{-j} \, , \hspace{0.3cm} W_{-j} 
= \sum_{i\neq 1}\frac{|V_{i,-j}|^2}{\epsilon_{-j} -\epsilon_i}
\label{6}
\end{equation}
Again the summation over $i$ $(j)$ is for positive (negative)
energy states. The intermediate state of $i=1$ is excluded
because it is already occupied. Then the total energy shift of
the system, including the vacuum background, is given by
\begin{equation}
W = W_1 + W_{\rm vac} \,  .
\label{7}
\end{equation}
The $W_1$ and $W_{\rm vac}$ are both negative and hence $W$ is
negative. Note that $W_{\rm QM}$ can be positive because of the
contributions from the negative energy intermediate states.

{\em Method II}. In this method we disregard the Pauli principle 
in all intermediate states. We can rewrite the above $W$ by 
adding and subtracting the same terms as
\begin{equation}
W' = W'_1 + W'_{\rm vac} , \hspace{0.3 cm} W'_1 = W_{\rm QM} \, ,
\label{8}
\end{equation}
\begin{equation}
W'_{\rm vac} = \sum_{j}W'_{-j} \, , \hspace{0.2 cm} W'_{-j} =
\sum_{i}\frac{|V_{i,-j}|^2}{\epsilon_{-j} -\epsilon_i} + \sum_{k
\neq j}\frac{|V_{-k,-j}|^2}{\epsilon_{-j} -\epsilon_{-k}} \, ,
\label{9}
\end{equation}
where $|-k\rangle$'s are also negative energy unperturbed states
\cite{6,7}.  The restriction $i\neq 1$ has been removed in the
$i$-summation for $W'_{-j}$.  Each of $W'_1$ and $W'_{-j}$ is
the energy shift in the context of single-particle quantum
mechanics. The $W'_{\rm vac}$ is the vacuum energy {\em in the 
absence of the particle in $|1\rangle $}. The $W'_1$ and 
$W'_{\rm vac}$ both contain terms that violate the Pauli principle 
but such terms all cancel out when they are added to obtain $W'$. 
Note also that the effects of transitions between negative energy 
states cancel out,
\begin{equation}
\sum_{j}\sum_{k \neq j}\frac{|V_{-k,-j}|^2}{\epsilon_{-j}
-\epsilon_{-k}} = 0 \, .
\label{10}
\end{equation}
The formal equality between $W$ of Eq. (\ref {7}) and $W'$ of
Eq. (\ref {8}) illustrates Feynman's prescription~\cite{1}. Let us warn, 
however, that this equivalence relies on the convergence of the 
series involved, in particular, the $j$-summation of 
$W'_{\rm vac}$ of Eq. (\ref {9}) that involves Eq. (\ref{10}).

In quantum field theory no negative-energy particles appear but
antiparticles of positive energies appear instead. The
unperturbed state that we consider is ${c_1}^\dagger |{\rm
vac}\rangle$. Here $|{\rm vac}\rangle $ is the state that
contains neither particles nor antiparticles at all. The energy 
of this unperturbed vacuum is set to zero. The ${c_1}^\dagger $ 
is an operator that creates a particle with energy $\epsilon_1$ 
and wave function associated with $|1\rangle$. The $|{\rm
vac}\rangle$ and ${c_1}^\dagger |{\rm vac}\rangle$ are the
ground states of the unperturbed system within the zero-particle
and one-particle sectors, respectively. Note that the particle
number is a conserved quantity of the model under consideration.
The external electric field leads to creation of a
particle-antiparticle pair, and so on. 	In this way the whole 
language of hole theory can be transcribed into that of quantum 
field theory.

\section{One-dimensional bag model}
We explicitly illustrate what we have shown in Sec. II by means
of model calculations. Let us consider the one-dimensional bag
model \cite{8,9}, which is a relativistic version of the
infinite square-well potential model of nonrelativistic quantum
mechanics. We define the model by the Dirac equation in one
dimension,
\begin{equation}
H_0\psi(x) = [\alpha p + \beta m + \beta S(x)] \psi(x) =
\epsilon \psi(x) \, ,
\label{11}
\end{equation}
where $m$ is the mass of the particle, $S(x)$ is a Lorentz
scalar potential and $p = -i d/dx $. We use units such that $c =
\hbar = 1$.  For the $2\times 2$ Dirac matrices, we use $\alpha
= \sigma_y$ and $\beta = \sigma_z$, where $\sigma_y$ and
$\sigma_z$ are the usual Pauli matrices.  For $S(x)$, we assume
that
\begin{eqnarray}
S(x) = \left\{
\begin{array}{cll}  S_0 \;\;\; & {\rm for}\;\; & |x|>a \\
			    0\;\;\;  & {\rm for}\;\; & |x|<a
\end{array}
      \right. ,
\label{12}
\end{eqnarray} 
where $S_0$ is a positive constant. It is understood that we let
$S_0 \to \infty$. For the bag model in three dimensions as a model 
of hadrons, see Ref.\cite{8,9}.

With the specific choice of $\alpha$, no complex numbers appear
in the Dirac equation. We write the wave function $\psi(x)$ as
\begin{eqnarray}
\psi(x) = \left( \begin{array}{c}u(x) \\ v(x) \end{array} \right).
\label{13}
\end{eqnarray}
The $u(x)$ and $v(x)$ vanish outside the bag, i.e., for $|x|>a$,
and are discontinuous at $|x|= a$. They are subject to the
boundary condition
\begin{equation}
u(\pm a) = \mp v(\pm a) \, .
\label{14}
\end{equation}
The scalar density $\psi^{\dag}\beta\psi = u^2-v^2$ vanishes at
$|x|=a$, but the vector density $\psi^{\dag}\psi = u^2+v^2$ does
not have to vanish at $|x|=a$.

The solutions of Eq. (\ref{11}) can be classified in terms of
parity.  For even parity, we obtain
\begin{equation}
u(x) = N \cos kx \, , \; \; \;\; v(x) = - N \frac{k \sin
kx}{\epsilon + m}
\label{15}
\end{equation}
where $k=\sqrt{\epsilon^2 - m^2}$ and $N$ is a normalization
factor.  For negative parity, we similarly obtain
\begin{equation}
u(x) = N \sin kx \, , \;\; \;\; v(x) 
= N \frac{k \cos kx}{\epsilon + m} \, .
\label{16}
\end{equation} 
Equation (\ref{14}) leads to
\begin{equation}
\tan ka = \frac{\pm \epsilon + m}{k} \, ,
\label{17}
\end{equation}
Where the double sign is $+$ $(-)$ for positive (negative)
parity.  Equation (\ref{17}) determines $k_n$ and $\epsilon_n$
(with $n = 0, 1, 2, ...$) for each parity. The $\epsilon_n$ can
be positive or negative. The normalization factor $N$ is given
by
\begin{equation}
N^2 = \frac{\epsilon(\epsilon +m)}{m+2a\epsilon^2} \, ,
\label{18}
\end{equation}
which applies to both of Eqs. (\ref{15}) and (\ref{16}). When
the potential for the Dirac equation is a pure Lorentz scalar, 
there is symmetry between positive and negative energies. This
symmetry is manifest in our model. For a positive parity state
with energy $\epsilon$, there exists a negative parity state of
energy $-\epsilon$. This can be seen through Eq. (\ref{17}).

The special case of $m=0$ is very simple and instructive. In
this case the solutions $k_n$ (with $n =0,1,2, ...$) of Eq. 
(\ref{17}) are given by
\begin{eqnarray}
k_n &=& \left(n+\frac{1}{4}\right)
\frac{\pi}{a} \;\; {\rm for}\;
\left\{ \begin{array}{ll}
{\rm even\, parity, }\;\;& \epsilon >0\, , \\ 
{\rm odd\, parity, } \;\;& \epsilon <0\, , 
\end{array}\right. \nonumber \\ 
&\label{19}\\
k_n &=& \left(n+\frac{3}{4}\right)
\frac{\pi}{a} \;\; {\rm for} \;
\left\{ \begin{array}{ll}
{\rm even\, parity, } & \epsilon <0\, , \\ 
{\rm odd\, parity, } & \epsilon >0\, ,
\end{array}\right. \nonumber 
\end{eqnarray}

In Sec. II we designated the energy levels with $i$, $-j$
and $-k$. For the bag model, however, we denote the levels with
$n^{ps}$ where $p$ stands for parity, $s$ is the sign of the 
energy and $n=0,1,2, \, ...$. For example $0^{++}$ and $0^{-+}$ 
are the lowest and the second lowest positive energy states,
respectively. They are $|1 \rangle$ and $|2 \rangle$, 
respectively, in the notation of Sec. II. If we denote the 
energy of state $n^{ps}$ with $\epsilon (n^{ps})$ we obtain
\begin{eqnarray}
\begin{array}{lrlr}
\epsilon (n^{++}) =& {\displaystyle 
\left( n+\frac{1}{4}\right) 
\frac{\pi}{a} } \, , \;\; 
&\epsilon (n^{-+}) =& {\displaystyle 
\left(n+\frac{3}{4} \right) 
\frac{\pi}{a} } \, ,
\nonumber \\ 
\epsilon (n^{+-}) =& {\displaystyle 
-\left( n+\frac{3}{4}\right) 
\frac{\pi}{a} } \, , \;\;
&\epsilon (n^{--}) =& {\displaystyle 
-\left( n+\frac{1}{4}\right)
\frac{\pi}{a} } \, .
\end{array} 
\label{19a}\\
\end{eqnarray}
Figure 2 shows the energy spectrum of this case of $m=0$. The
energy levels are all equally spaced. If mass $m$ is increased
from 0, the energy levels are pushed away from $E=0$, the levels
nearer to $E=0$ being more affected than those further away from
$E=0$.

\vskip -1cm
\vskip 1cm
\begin{figure}[htb]
{\psfig{figure=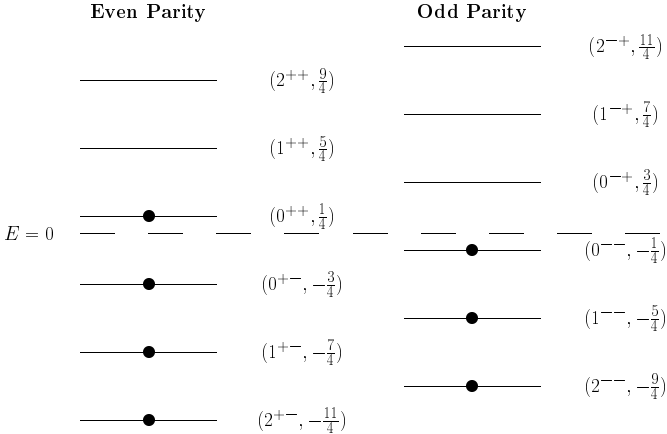,width=2.5in}}
\vskip -1.5cm
\caption{
Energy levels of the bag model with $m=0$.
The pair of numbers in brackets are respectively quantum number 
$n^{ps}$ and $\epsilon$ in units of $\pi/a$. 
The dots indicate occupied levels.
}\label{fig2}
\end{figure}

Let us now assume the external perturbation
\begin{equation} 
V(x) = \lambda x \, ,
\label{20}
\end{equation}
and work out the second order energy shifts $W$ and $W'$
explicitly for the bag model. If the charge of the particle is
$q$ and the external perturbation is due to a constant electric
field ${\cal E}$ along the $x$-axis, then $\lambda = -q{\cal
E}$. Then the second order energy shift $W$ takes the form
\begin{equation}
W = -\frac{1}{2}P{\cal E}^2 \, .
\label{21}
\end{equation}
The coefficient $P$ defines the electric polarizability of the
system. The $W'$ can be written similarly with $P'$.

Interaction $V(x)$ connects states of opposite parity. The matrix 
element between states $n^{+s}$ and ${n'}^{-s'}$ is given by
\begin{eqnarray}
&&\langle {n'}^{-s'} |V| n^{+s} \rangle \equiv
V_{{n'}^{-s'}, n^{+s}} = 
\nonumber \\
&&\lambda N N'\int^a_{-a}\left[\cos kx \sin k'x
- {\large \frac{kk'\sin kx \cos k'x}{(\epsilon +m)(\epsilon ' +m)}}
\right]x dx \, ,
\label{22}
\end{eqnarray}
where $k$ and $k'$ are associated with the states of $n$ and
$n'$, respectively, and similarly for $N, \ N', \ \epsilon$ and
$\epsilon '$.  Let us first examine the simple case of $m = 0$,
which gives us much insight into the problem. Then the matrix
element becomes
\begin{equation}
|\langle {n'}^{-s'} |V| n^{+s} \rangle| =
\frac {\lambda}{a(\epsilon - \epsilon ^\prime )^2 } \, .
\label{23}
\end{equation}
For the quantities of Eqs. (\ref{5}-\ref{7}) of method I we obtain
\begin{equation}
W = W_{0^{++}}+ W_{\rm vac} \, ,
\label{24} 
\end{equation}
\begin{equation}
W_{0^{++}} = \sum_{n'=0}^{\infty} f(n'+{\small \frac {1}{2}}) \, ,
\label{25}
\end{equation}
\begin{equation}
W_{\rm vac} = \sum_{n=0}^{\infty} (W_{n^{+-}}+ W_{n^{--}}) \, ,
\label{26}
\end{equation}
\begin{equation}
W_{n^{+-}} = \sum_{n'=0}^{\infty} f(n+n'+{\small \frac{3}{2}}) \, ,
\label{27}
\end{equation}
\begin{equation}
W_{n^{--}} = \sum_{n'=1}^{\infty} f(n+n'+{\small \frac{1}{2}}) 
=W_{n^{+-}} \, ,
\label{28}
\end{equation}
where
\begin{equation}
f(x) = -\frac {\lambda^2 a^3}{\pi^5}\frac{1}{x^5} \, . 
\label{29}
\end{equation}
In Eq. (\ref {25}) the term with the argument $n'+\frac{1}{2}$ 
is due to the transition $0^{++} \to {n'}^{-+}$. 
In Eq. (\ref{27}) the term with  $n+n'+\frac {3}{2}$ is due
to $n^{+-}\to {n'}^{-+}$. Note that $n'=0$ is excluded in the 
summation for $W_{n^{--}}$. The above series all converge very 
rapidly. In Eq. (\ref{25}) the first term with $n'=0$ constitutes 
99\% of the sum.

Next, let us turn to the $W'$ of method II, Eq. (\ref{8}).  Let
us again consider the $m=0$ case. Curiously enough, it turns out
that the energy shifts of the individual levels all vanish and 
consequently the total energy shift is zero in this 
case, i.e., in the notation of Sec. II,
\begin{equation}
W'_1= W'_{-j} = 0 \, , \hspace{0.3cm} W'=0 \, .
\label{30}
\end{equation}
For example, we find that $W'_ {0^{++}}$ ($=W'_1$) is of the 
structure
\begin{equation}
W'_ {0^{++}} = \sum_{n'=0}^{\infty} 
\left[f(n'+{\small \frac {1}{2}}) + 
f(-n'-{\small \frac {1}{2}})\right] = 0 \, .
\label{31}
\end{equation}
Recall that $f(x)$ is an odd function. The two terms in the square 
brackets respectively correspond to the two terms of the right 
hand side of $W_{\rm QM}$($=W'_1$) of Eq. (\ref{3}). The first 
(second) term is due to the intermediate states of positive 
(negative) energies. Exactly the same situation is found for the 
energy shift of each of other states, that is,
\begin{equation}
W'_{n^{+-}}
= \sum_{n'=0}^{\infty}\left[f(n+n'+{\small \frac{3}{2}}) + 
f(n-n'+{\small \frac{1}{2}})\right] = 0 \, ,
\label{32}
\end{equation}
\begin{equation}
W'_{n^{--}}
= \sum_{n'=0}^{\infty}\left[f(n+n'+{\small \frac{1}{2}}) + 
f(n-n'-{\small \frac{1}{2}})\right] = 0 \, .
\label{33}
\end{equation}
The vanishing of $W'_{n^{+-}} $ and $ W'_{n^{--}} $ given above 
may not be immediately obvious. In Appendix we show that the 
series of Eqs. (\ref{32}) and (\ref{33}) can be rewritten 
exactly in the form of Eq. (\ref{31}). The result of $W'=0$ is in 
clear contradiction with $W$ of Eq. (\ref{24}) that is nonzero 
and negative. This is very puzzling. The vanishing of $W'$ 
means that the system is rigid against the external perturbation, 
which we find intuitively strange.

In Sec. II we warned that the equivalence between $W$ and $W'$
relies on the assumption that the summations involved converge.
What happens in the above puzzle is the following. Let us first
explain it by using the notation of Sec. II. There is no problem
in convergence of the summations except for the $j$ summation
of Eq. (\ref{9}). Each of $W'_{-j}$ is well defined, but when it
is summed with respect to $j$, we obtain the double sum of the 
left hand side of Eq. (\ref{10}). As can be seen from Eq. 
(\ref{23}), $|V_{-k,-j}|^2$ depends on $j$ and $k$ only through 
the difference $j-k$. Coming back to the notation of this section, 
the left hand side of Eq. (\ref{10}) becomes
\begin{equation}
\sum_{n=0}^{\infty} \sum_{n'=0}^{\infty}
\left[f(n-n'+{\small \frac{1}{2}}) +
f(n-n'-{\small \frac{1}{2}})\right] \, .
\label{34}
\end{equation}
The summations with respect to $n$ and $n'$ individually converge.
When the two summations are combined, however, we realize that
Eq. (\ref{34}) involves something like the alternating series
$1-1+1-1+1 \, ...$ . We show this explicitly in Appendix.
This series can converge only conditionally at best.
Its sum depends on how the series is arranged. Equation (\ref{34})
was set to zero in rewriting $W$ into $W'$; see Eq. (\ref{10}). 
In the way as $W'$ is explicitly worked out as shown above in
method II, that is, the $n'$ summation is done {\em before} the 
$n$ summation, the series of Eq. (\ref{34}) is actually arranged 
such that its sum assumes a nonzero value. This is where the 
discrepancy between methods I and II stems from. Let us emphasize 
that the conspiracy of the above alternating series is well 
hidden in the sense that all the (single) series that appear in 
the steps of method II are absolutely convergent.

We have also examined the case with nonzero values of mass $m$. 
The calculation is lengthy but straightforward. So we do not 
describe it. We have confirmed that essentially the same situation 
persists, that is, the results of the two methods disagree. For 
the states of very large values of $j$ and/or $k$, effects of the 
finite mass $m$ becomes negligible. Therefore the nonconvergence 
aspect of the series involved is not essentially affected by $m$. 
Figure 3 shows $W$ (solid line) and $W'$ (dashed line) as 
functions of $ma$. Note that the difference between the two is 
larger for smaller $ma$. Figure 3 also shows the nonrelativistic 
limit (dotted line) that we derive in Sec. IV.

\vskip -0.5cm
\begin{figure}[htb]
\centerline{\psfig{figure=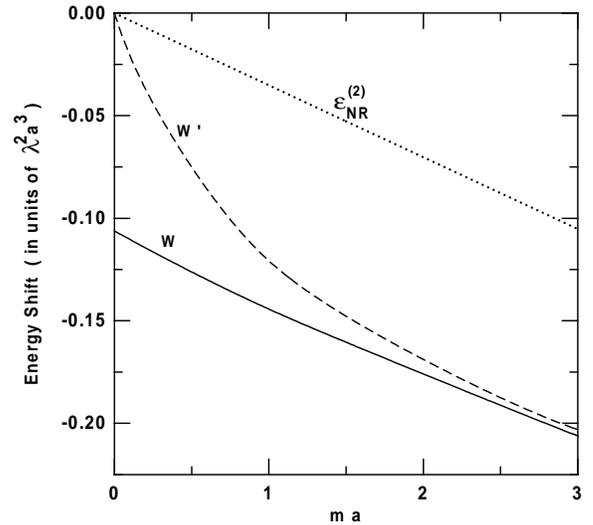,width=3.in}}
\vskip -1cm
\caption{
Relativistic energy shifts $W$ of method I (solid line), 
$W'$ of method II (dashed line) and their nonrelativistic counterpart 
${\epsilon_{\rm NR}}^{(2)} $ of Eq. (\ref{42}) (dotted line) are 
shown in units of $\lambda ^2 a^3$ as functions of $ma$. 
The ${\epsilon_{\rm NR}}^{(2)} $ is not meaningful unless $ma\gg 1$.
}\label{fig3}
\end{figure}

\section{The Dalgarno-Lewis method}
The calculation of method II that was presented in Secs. II and 
III is somewhat involved. So it would be good to confirm it by 
repeating the calculation in a different manner. We do so by
using the DL method. The DL method is an alternative form of 
perturbation theory in which summations over intermediate states 
are avoided \cite{2,3,4}. As a price for it, one has to solve an
inhomogeneous differential equation. The DL method is often used
in calculating the electric polarizability of nonrelativistic
bound systems. There is another similar, powerful method called
logarithmic perturbation expansion \cite{10,11}, which we do not
use here. Consider any one of the energy levels. Let its
unperturbed wave functions be $\psi^{(0)}(x)$ and its first
order perturbation be $\psi^{(1)}(x)$. The $\psi^{(1)}(x)$ can
be determined by the DL equation
\begin{equation}
(H_0-\epsilon^{(0)})\psi^{(1)}(x) = - V(x)\psi^{(0)}(x) \, ,
\label{35}
\end{equation}
where $\epsilon^{(0)}$ is the unperturbed energy, i.e.,
$\epsilon$ of Sec. III.  We write $\psi^{(1)}$ as
\begin{eqnarray}
\psi^{(1)}(x) 
= \left( 
\begin{array}{c}u^{(1)}(x) \\ 
v^{(1)}(x) 
\end{array}
\right).  
\label{36}
\end{eqnarray}
Its components are again subject to boundary condition
(\ref{14}),
\begin{equation}
v^{(1)}(a) = -u^{(1)}(a) \, .
\label{37}
\end{equation}
When $\psi^{(1)}$ is found, the second order energy shift
$\epsilon^{(2)}$ can be calculated by
\begin{equation}
\epsilon^{(2)} 
= \int^\infty_{-\infty}V(x)\psi^{\dag (1)}(x)\psi^{(0)}(x)dx \, .
\label{38}
\end{equation}
The summation over intermediate states is done implicitly. It is 
understood that there is no restriction on intermediate states
due to the Pauli principle.  Since it includes all intermediate 
states of negative as well as positive energies, $\epsilon^{(2)}$ 
is nothing but $W'_1$ or $W'_{-j}$ of Sec. II. If we take the
unperturbed wave function for $|1\rangle$ for $|\psi ^{(0)} (x)$, 
for example, we obtain $W'_1$. Obviously the DL method is not 
useful for method I.

By solving Eq. (\ref{35}) we obtain
\begin{eqnarray}
u^{(1)}(x) = \frac{\lambda N}{2k^2}\left[mx\cos kx +
\eta \epsilon k(x^2-a^2)\sin kx 
\right.\nonumber\\ \left.
-\frac{\eta mk}{2\epsilon}
\left(\frac{1}{\epsilon +m}+2a\right)\sin kx
\right] \, ,
\label{39}
\end{eqnarray}
\begin{eqnarray}
v^{(1)}(x) = \frac{\lambda m N}{2\epsilon k^2}
\left\{ \frac{\eta \epsilon}{\epsilon +m}kx\sin kx +
\left[\frac{1}{2}-(\epsilon -m)a
\right.\right.\nonumber\\ \left.\left.
+\frac{\epsilon^2(\epsilon -m)}{m}
(x^2-a^2)\right] \cos kx \right\},
\label{40}
\end{eqnarray}
where $\eta = 1 (-1)$ for even (odd) parity.  The $\epsilon$ and
$k$ are $\epsilon^{(0)}$ and $k^{(0)}$, respectively.  With
these $u^{(1)}$ and $v^{(1)}$ in Eq. (\ref{38}) we arrive at 
\begin{eqnarray}
\epsilon^{(2)} &=&{\large \frac{\lambda ^2 m}
{24 k^4\epsilon a [2(\epsilon a)^2+ma]}} \nonumber\ \\ 
&\times& \left\{ 2(ka)^2(ma+3) [4(\epsilon a)^2-6ma-3]
\right. \nonumber \\ && \left. 
-15(ma)^2(2ma+1) \right\} \, .
\label{41}
\end{eqnarray}
This applies to any of the energy levels with an appropriate choice 
of $k$ that is subject to Eq. (\ref{17}). We have explicitly 
confirmed that $\epsilon^{(2)}$ agrees with $W'_1$ or $W'_{-j}$ of 
method II for each of the energy levels and hence the same $W'$ as
that of Sec. III. The $\epsilon^{(2)}$ vanishes if $m=0$. This is 
consistent with what we found in Sec. III.

In the non-relativistic limit of $m \to \infty$, Eq. (\ref{37}) 
is reduced to
\begin{equation}
{\epsilon_{\rm NR}}^{(2)} = \frac{\lambda^2 m}{24 k^4}
\left[4(ka)^2 - 15 \right] \, , \hspace{.4 cm} 
ka = \frac {\pi}{2} \, .
\label{42}
\end{equation}
The nonrelativistic value of $ka$ follows from Eq. (\ref{17})
with the plus sign and $m \to \infty$. Equation (\ref{38}) agrees 
with the result for an infinite square-well potential of 
non-relativistic quantum mechanics \cite{12}. The 
${\epsilon_{\rm NR}}^{(2)} $ is compared with its relativistic 
counterparts $W$ and $W'$ in Fig. 3. Note that even when $m$ is 
as large as $m =3/a$ (or $m\approx 600$ MeV if $a=1$ fm), 
the relativistic energy shifts are about twice as large as their 
nonrelativistic counterpart.

\section{summary and discussions}
For a system consisting of a particle bound in a given potential
together with its vacuum background, we examined the second
order energy shift caused by external perturbation $V$. We
examined two formally equivalent methods of calculation, I and
II. Method I takes account of the Pauli principle in
intermediate states whenever it is applicable. In method II the
Pauli principle is completely ignored. We showed that, if the
energy shifts of all occupied levels are summed up in method II,
the terms violating the Pauli principle formally cancel out.
Thus the two methods appear equivalent. This illustrates
Feynman's prescription.

This equivalence, however, is not free from ambiguity. We
calculated the energy shift explicitly for the one-dimensional
bag model with external perturbation $V(x)=\lambda x$. As shown 
in Fig. 3, the two methods lead to different energy shifts.
Thus Feynman's prescription fails in this example. For method II, 
we did the calculation in two different manners, one by summing 
up over the intermediate states and the other by using the DL
method. The same result were obtained by two calculations. In
method II the energy shifts of the individual occupied levels 
are unambiguously obtained. When they are summed over all 
negative energy states, however, an alternate series emerges.
The sum of the series depends on how the summation is done. This 
is essentially the source of the discrepancy between the two
apparently equivalent methods. The alternating series is hidden 
such that, if one simply follows method II, one would not notice 
it.

Feynman's prescription fails in the specific example that we 
have described. A question naturally arises here. Does similar 
difficulty arise in more general situations? We suspect that it 
may well do. Let us first point out that, although we assumed 
a specific form of external perturbation $V(x) = \lambda x$, 
Feynman's prescription fails in the one-dimensional bag model 
irrespectively of the form of $V(x)$. Again for simplicity let 
us assume $m=0$. Then the matrix element of $V(x)$ are of the 
form of
\begin{equation}
\int^a_{-a} V(x)\sin [(k-k')x] dx \;\;\;\; {\rm or} \;\;\;\;
\int^a_{-a} V(x)\cos [(k-k')x] dx \, .
\label{43}
\end{equation}
No matter how large $k$ and $k'$ become, the matrix element 
is of the same form as that of Eq. (\ref{23}). This feature 
remains essentially the same when $m$ becomes nonzero. In this
connection, recall what we said in the last paragraph of Sec. 
III.

Next let us consider the three-dimensional bag model, subject 
to a constant external electric field. The perturbation 
interaction can be taken as $\lambda z = 
\lambda r \cos \theta $. For states with large quantum numbers, 
the radial part of the wave function is similar to the 
one-dimensional wave function. This is so in the sense that at 
large distances the spherical Bessel functions involved are like 
the sine and cosine functions. For the angular part, the matrix 
element of $\cos\theta $ between two adjacent angular momentum 
states has a part that remains finite no matter how large the 
angular momenta become. Therefore, the alternating series 
involved in method II will remain.  We are aware of a few 
calculations of the electric and magnetic polarizabilities of 
the nucleon by using the bag model \cite{13,14,15}. Method I 
was used in these calculations and hence the problem with 
method II was not encountered.

We have assumed that the energy spectrum of the unperturbed 
system is discrete. The case of continuum spectrum can be 
handled by enclosing the system in a very large cavity. The
unperturbed Hamiltonian $H_0$ in this case can be that of
the bag model (with a large radius) plus some other 
interaction that produces states localized, say, around the
origin. Let us consider such a model in one dimension. The 
perturbation of the form of $V(x)=\lambda x$, if taken 
literally, would not make much sense because it becomes very 
large as $x $ approaches the cavity radius. If one chooses 
$V(x)$ such that it remains reasonably small within the entire 
cavity, one can treat it by perturbation theory. Then the 
calculation will go in a way essential the same as we have 
done. Feynman's prescription will probably fail again.

As far as we know the example that we have presented is the 
first counter-example against Feynman's prescription that 
seems to have been taken for granted for many years. If we 
have to choose between methods I and II, we are inclined to 
take method I and abandon method II that is based on Feynman's 
prescription. We think that, if we encounter ambiguity by 
disregarding the Pauli principle, we should remain faithful 
to the Pauli principle in every step of calculation. In view 
of the fact that Feynman's prescription has been used 
extensively, its possible failure may have serious implications.

\section*{Acknowledgements}
This work was supported by Funda\c c\~ao de Amparo \`a Pesquisa
do Estado de S\~ao Paulo (FAPESP), Conselho Nacional de
Desenvolvimento Cient\'\i fico e Tecnol\'ogico (CNPq) and the
Natural Sciences and Engineering Research Council of Canada. YN
is grateful to Universidade de S\~ao Paulo and Instituto de 
F\'\i sica Te\'orica of Universidade Estadual Paulista for warm 
hospitality extended to him during his visits of 1996 and 1998.

\vskip 1cm
\begin{center}
{\bf Appendix}
\end{center}
Let us first examine how $W'_{n^{+-}}$ of Eq. (\ref{32}) and  
$W'_{n^{--}}$ of Eq. (\ref{33}) vanish. First note that
\begin{equation}
W'_{n^{+-}} = W'_{(n+1)^{--}} \, .
\label{A1}
\end{equation}
Therefore it is sufficient to show that $W'_{n^{--}} = 0$.
This can be seen as follows:
\begin{eqnarray}
W'_{n^{--}} =&& \sum_{n'=0}^{\infty}f(n+n'+{\small \frac{1}{2}})
 + \sum_{n'=0}^{n-1}f(n-n'-{\small \frac{1}{2}}) \nonumber \\  
 &&+  \sum_{n'=n}^{\infty}f(n-n'-{\small \frac{1}{2}}) \, .
\label{A2}
\end{eqnarray}
If we define $\nu=n'-n$, the last sum can be reduced to
\begin{equation}
\sum_{\nu=0}^{\infty}f(-\nu - {\small \frac{1}{2}}) \, .
\label{A3}
\end{equation}
It is not difficult to see that the first two sums can be 
combined into
\begin{equation}
\sum_{\nu=0}^{\infty}f(\nu +{\small \frac{1}{2}}) \, .
\label{A4}
\end{equation}
\newpage
Because of $f(x)=-f(-x)$, $W'_{n^{--}} = 0$ follows.
Although $W'_{n^{--}}$ can be regarded as an alternating series,
it is absolutely convergent. It is not like the alternating 
series that we mention below Eq. (\ref{34}).

Next let us examine the structure of the double series of
Eq. (\ref{34}). Consider a set of $(n,n')$ such that $n=n'+1$,
i.e.,
\begin{equation}
(n,n') = (1,0),\, (2,1),\, (3,2),\, \cdots \, .
\label{A5}
\end{equation}
For this set we find that the term in the square brackets of Eq. 
(\ref{34}) takes the same value $f(\frac{3}{2})+f(\frac{1}{2})$. 
This is so no matter how large $n$ and $n'$ individually are. 
Similarly, for a set of $(n,n')$ such that $n=n'-1$, i.e.,
\begin{equation}
(n,n') = (0,1),\, (1,2),\, (2,3),\, \cdots \, ,
\label{A6}
\end{equation} 
we find $f(-\frac{1}{2})+f(-\frac{3}{2})
= -[f(\frac{3}{2})+f(\frac{1}{2})]$. Therefore, the
terms corresponding to the sets of $(n,n'=n \pm 1)$ can be
seen as an alternating series like $1-1+1-1+1 \cdots$. 
If we pair the above like $(1,0)$ and $(0,1)$, $(2,1)$ and 
$(1,2)$, $\cdots$, then we find the double sum vanishes, 
like Eq. (\ref{10}). If we pair the above like $(1,0)$ and
$(1,2)$, $(2,1)$ and $(2,3)$, $\cdots$, then the sum does not
vanish. We find similar series for $(n,n'=n \pm 2)$,
$(n,n'=n \pm 3)$, and so on. This shows that the sum of the
double series has an ambiguity that is related to how the
$n$-$n'$ summation is done.


\begin{references}
\bibitem{5} We can start with a negative energy bound state,
e.g., $|-1\rangle$. If we do so, however, we will meet
unnecessary, nonessential complications in the context of
quantum field theory.
\end{references}
\end{document}